\documentclass[aps,pra,twocolumn,groupedaddress]{revtex4-2}  
\usepackage{graphicx}
\usepackage{mathtools}
\usepackage{amssymb}
\usepackage{amsmath}
\usepackage{newtxtext,newtxmath,lipsum}
\usepackage{xcolor}
\usepackage[shortlabels]{enumitem}

\begin{document}
\bibliographystyle{revtex}

	\title{Second harmonic phase locking and synchronization blockade in quadratically coupled driven quantum van der Pol oscillators }

	\author{Nissi Thomas}
	\affiliation{Department of Nonlinear Dynamics, School of Physics, Bharathidasan University \\ 
		Tiruchirappalli - 620024, Tamil Nadu, India.}
	
	\author{M. Senthilvelan}
	\affiliation{Department of Nonlinear Dynamics, School of Physics, Bharathidasan University \\ 
		Tiruchirappalli - 620024, Tamil Nadu, India.}

	\begin{abstract}
			 We investigate the  dynamics of a quadratically coupled system under the influence of an external drive applied to the second oscillator, where the coupling facilitates a high-order synchronization with phase-locking emerging in the form of $2:1$ between the oscillators. Our analysis reveals a synchronization blockade in the first oscillator, characterized by the complete suppression of conventional $1:1$ phase-locking with the drive. Instead the first oscillator exhibits a  $2:1$ phase-locking behavior arising from the anharmonicity  induced by the quadratic coupling. In contrast, we observe that the directly driven second oscillator synchronizes with the drive, showing $1:1$ phase-locking but notably at second harmonic frequency ($\omega_2=2\omega_d$). A classical mean-field analysis of the corresponding equations of motion reproduces this asymmetric phase-locking geometry, including the two symmetry-related 2:1 phase-locked states of the first oscillator and the 1:1 phase-locked state of the second oscillator. This demonstrates that the phase-locking structure itself can be understood from the nonlinear classical dynamics. The quantum analysis, however, reveals the microscopic origin of the synchronization blockade: the quadratic interaction imposes a two-phonon selection rule that suppresses the conventional single-phonon synchronization channel of the first oscillator. Furthermore, we show that the system exhibits mutual synchronization when both the oscillators satisfies the resonance condition, enabling coherent energy exchange facilitated by nonlinear quadratic coupling. The mutual synchronization shows synchronized regimes and also subtle suppression of synchronized regimes near resonance occurring due to spectral splitting of the energy states. Using perturbation analysis of the master equation within the low excitation subspace, we analyze steady-state phase distribution and synchronization measures, supported by population statistics and spectral responses. We also propose possible experimental realizations in trapped-ions and optomechanical setups. These findings highlight the crucial role of quadratic coupling in enabling nonclassical synchronization phenomena, offering deeper insights for quantum control strategies and the development of quantum information platforms.
	\end{abstract}

\maketitle
\section{INTRODUCTION}\label{intro}
Synchronization is a fundamental collective phenomenon where self-sustained oscillators with inherent frequency mismatches align their rhythms through coupling, so that their phases or frequencies become correlated \cite{pikovsky}. In classical nonlinear dynamics this behavior appears across electronics \cite{behta, ignatov}, chemistry \cite{nakata, liu} and biology \cite{levy, gholizade} and exhibits not only $1:1$ phase-locking but also high-order phase-locking where oscillator synchronize at rational frequency ratios. High-order synchronization has been explored in different contexts, ranging from coupled electromechanical oscillators \cite{montoya}, coupled van der Pol oscillators \cite{nis}, Josephson junctions \cite{valkering} and neuron models \cite{abhay}. Advances in microscopic models \cite{amitai, weiss, fan,  liao, qiao, matheny, hriscu, rodrigues, tilley, roulet, roulet2, laskar, tindall, xu, zhu, cabot} has extended the idea of synchronization in quantum realm, where conventional $1:1$ phase-locking and high-order locking \cite{nissi} can occur between the quantum oscillators and external drives. At the same time, the quantum synchronization exhibits some genuine quantum behaviors such as quantized phase-locking \cite{lorch}, nonclassical correlations \cite{lee2, sonar, ameri}, quantum synchronization blockade \cite{lorch2} and so on.

Quantum synchronization blockade is a unique quantum phenomenon where synchronization between coupled quantum oscillators is suppressed-not by classical noise or detuning, but by quantum interference effects and system symmetries, leading to regimes of bistable or higher-order phase distributions instead of conventional $1:1$ locking. At its core, synchronization blockade arises from interferometric cancellation: destructive quantum interference between different transition pathways prevents the oscillators from locking their phases and frequencies in the usual way \cite{roulet, roulet2, koppenhofer, tan, kehrer1}. This effect is closely tied to the symmetries of the system's Hamiltonian and the structure of its quantum states \cite{solanki}. Also the mechanisms like few level anharmonicity \cite{lorch2}, non-reciprocal couplings \cite{kehrer2} that engineer directional energy flow between the oscillators can realize this behavior. In our recent work we investigated the synchronization blockade induced by normal mode splitting in a linearly coupled (reactive coupling) quantum van der Pol oscillator under the influence of external drive \cite{nis2}. In all these settings coupling between the oscillators is linear. This naturally raises the question of how quadratic coupling reshapes quantum synchronization blockade and high-order phase locking, especially under asymmetric driving, where only one oscillator couples directly to the external drive. 

Quadratic coupling introduces fundamentally different physics by enabling nonlinear phonon exchange that are absent in linear regime. Unlike Kerr-type self nonlinearities that require challenging requirement on experimental setup quadratic interaction emerge in membrane-in-the-middle optomechanical setup \cite{xie} and trapped-ion platforms \cite{ding17}. By inducing the shift in energy spectrum which is directly dependent on the phonon number, it enables creation of nonclassical mechanical states \cite{nunnenkamp,htan}, facilitates optomechanically induced transparency \cite{huang, zhan} and opacity \cite{si}. Furthermore, it provides a physical path to achieve single-photon nonlinearities \cite{liao1, liao2} and the photon/phonon blockade effect \cite{xie, xwxu}.  

In this paper, upon considering quadratically coupled dissipative quantum van der Pol oscillators with selective driving of the second mode, we uncover asymmetric synchronization phenomena. Here we consider the quadratic coupling, where nature of the coupling introduces strong nonlinear interactions between the oscillators, even under intense driving conditions. When the second oscillator is driven, the interplay of the drive and nonlinear coupling induces significant anharmonicity in the system's energy spectrum-a key ingredient for realizing quantum synchronization blockade. Through perturbative analysis of the phase distribution, we show that the first oscillator is prevented from achieving $1:1$ phase locking with the drive and instead exhibits $2:1$ phase locking with bistability in the phase distribution, a direct consequence of the quadratic coupling-induced anharmonicity. In contrast, the second, directly driven oscillator synchronizes in a standard 1:1 manner with the external field but at the second harmonic frequency. Nonlinear energy exchange between the oscillators gives rise to mutual synchronization, resulting in a distinct phase correlation between the modes. By examining phonon statistics and employing synchronization measures, we elucidate how population transfer and energy level anharmonicity supports the emergence of synchronization blockade in one oscillator, while simultaneously facilitating synchronization in the other. To further clarify the origin of the asymmetric synchronization, we analyze the corresponding classical equations of motion obtained within the mean-field approximation. The resulting amplitude and phase equations show that the quadratic coupling itself produces a two-fold phase preference for the first oscillator and a single phase-locked state for the directly driven second oscillator. However, the classical equations do not by themselves reveal the microscopic origin of the suppression of conventional 1 : 1 synchronization in the first oscillator. The quantum analysis provides this microscopic picture through the two-phonon selection rule associated with the quadratic interaction, which suppresses the direct single-phonon synchronization channel and favors the higher-order 2 : 1 synchronization pathway. This comparison therefore distinguishes the classical nonlinear origin of the asymmetric phase-locking structure from the genuinely quantum mechanism underlying the synchronization blockade.

We organize our presentation as follows: In Sec. \ref{model}, we introduce the system model, detailing the quadratic interactions and external driving within the master equation framework for the van der Pol oscillator. In Sec. \ref{quant}, we analyze the phase-distribution to explore the resulting phase-locking dynamics in the quantum regime, emphasizing the role of nonlinear coupling. Here, we derive analytical expressions for the steady-state solutions of the master equation to obtain the phase distribution using perturbation analysis, highlighting quantum effects in phase-locking behavior. In Sec. \ref{synch_mea}, we define the synchronization measure and  investigate the synchronization mechanism by analyzing the phonon statistical properties of the system. We also present the characteristic properties of the system through power spectrum analysis in Sec. \ref{power_spe}. In Sec. \ref{exp_rea}, we discuss possible experimental realization of the system. Finally, in Sec. \ref{conc}, we present a summary and discussion of our main findings. In Appendix A we have provided classical steady-state analysis.

\section{Model}\label{model}

\par We consider quadratically coupled quantum van der Pol oscillator under the influence of external force. The master equation describing the dynamics of the system is given by \cite{nissi, lee}  
\begin{equation}
\dot \rho=-i[H_0+H_I,\rho]+\sum_{i=1}^{2}\gamma_1\mathcal{L}[a_i^{\dagger}]\rho+\gamma_2\mathcal{L}[a_i^2]\rho, \label{mas_eq}
\end{equation}
\begin{figure}
	\includegraphics[width=1.0\linewidth]{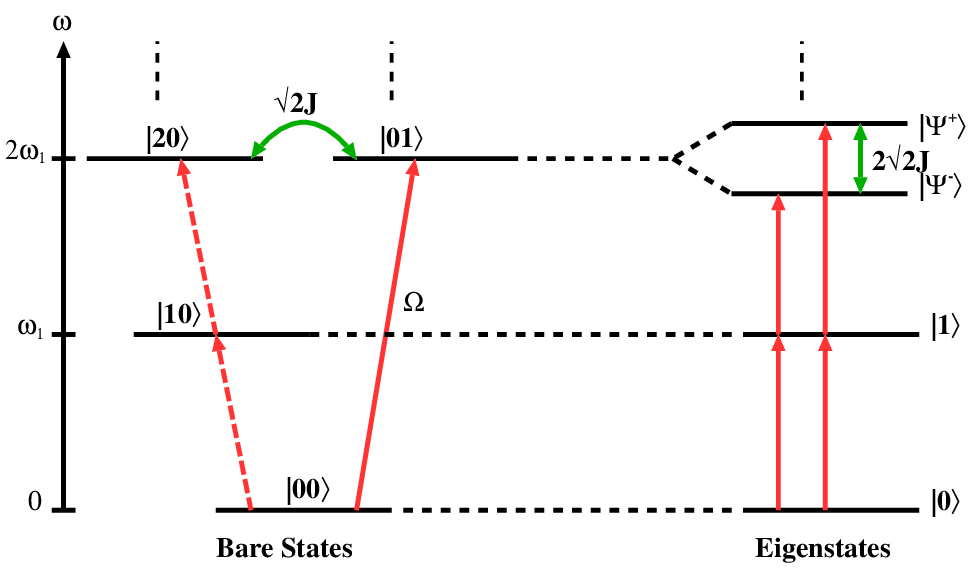}
	\caption{The Schematic energy level diagram of the quadratically coupled quantum van der Pol  oscillator under the influence of drive.}
	\label{energy}
\end{figure}
where $a_i (a_i^{\dagger})$ is the bosonic operator for $i^{th}$ oscillator with frequency $\omega_i$. The coefficients $\gamma_1$ and $\gamma_2$ are the linear and nonlinear damping rates which describes the non-unitary process, the system undergoes, with $\mathcal{L}[{\hat{x}}]\rho=\hat{x}\rho\hat{x}^{\dagger}-\frac{1}{2}\{\hat{x}^{\dagger}\hat{x}\rho+\rho\hat{x}^{\dagger}\hat{x}\}$. The Hamiltoninan $H_0=\sum_{i=1}^2\Delta_ia_i^{\dagger}a_i$ and the Hamiltonian, $H_I$ describes the interaction between the oscillator modes $a_i$ which comprises of quadratic coupling and an external force acting on the second oscillator mode.  In the rotating frame at the frequency of the drive $\omega_d$, the Hamiltonian $H_I$ can be described in the from
\begin{equation}
H_I = \Omega(a_2^{\dagger}+a_2)+J(a_1^{\dagger 2}a_2+a_1^2a_2^{\dagger}), \label{int_ham}
\end{equation}
where $\Omega$ is the drive strength,  $\Delta_1=\omega_1-\omega_d$ and $\Delta_2=\omega_2-2\omega_d$, are the detunings of the  first and second oscillator from fundamental and second-harmonic drive frequencies respectively and $J$ represents the quadratic coupling strength. The coupling term  mediates the exchange of two phonons in oscillator 1 with a single phonon in oscillator 2 such that the processes of the form: $|n_1n_2\rangle$ $\xleftrightarrow{}$ $|n_{1}+2,n_2-1\rangle$ occurs with effective amplitude $J\sqrt{(n_1+1)(n_1+2)n_2}$, where $n_1$ and $n_2$ are the phonon Fock states of first and second oscillator modes respectively. This coupling amplitude introduces anharmonicity into the energy spectrum as demonstrated in Fig. \ref{energy}. Throughout our analysis we focus in quantum regime with low excited states enforced by the condition $\gamma_2\gg\gamma_1$ together with weak driving $\Omega\le\gamma_1$. This limit of strong nonlinear damping $\gamma_2$ and weak driving confines the dynamics of the system (\ref{mas_eq}) predominantly to the subspace with a total excitation number $N=n_1+n_2\le2$, spanned by the basis states $N=0:\{|00\rangle\}$, $N=1:\{|10\rangle\}$ and $N=2:\{|20\rangle,|01\rangle\}$. The quadratic coupling $J$ does not act on the subspaces $N=0$ and $N=1$, but couples the bare states $|20\rangle$ and $|01\rangle$ within the second excitation ($N=2$) manifold. In the strong coupling regime these bare states becomes a pair of dressed states (non-degenerate states) $\Psi^{\pm}=(|20\rangle\pm|01\rangle)/\sqrt{2}$ as represented in Fig. \ref{energy} (right) with an energy splitting proportional to $J$. This few level anharmonicity in the energy spectrum is central to the synchronization properties we discuss in the following sections.
\section{Phase Distribution}\label{quant}
\begin{figure*}
	\includegraphics[width=0.85\linewidth]{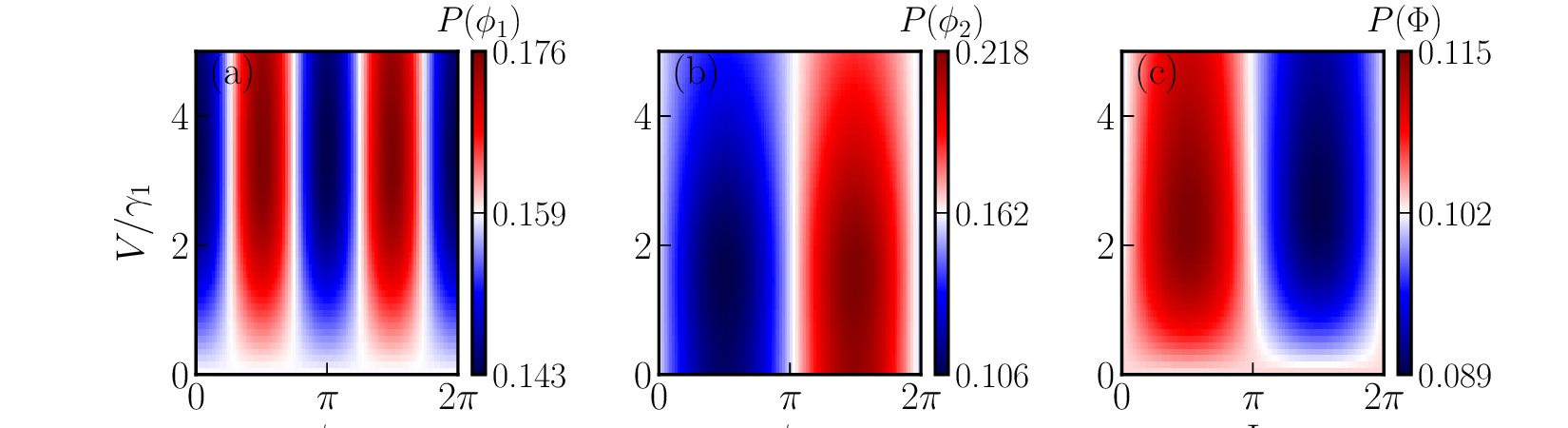}
	\caption{Phase distribution of the single oscillator modes (a) $P(\phi_1)$, (b) $P(\phi_2)$ and relative Phase distribution (c) $P(\Phi)$, defined in Eqs. (\ref{single_phase}) and (\ref{phase_relative}) respectively, as a function of coupling strengh $J$ with $\Omega/\gamma_1=0.5$, $\Delta_1/\gamma_1=\Delta_2/\gamma_1=0$ and $\gamma_2/\gamma_1=10$.}
	\label{phase_dist}
\end{figure*}
To characterize the phase-locking dynamics induced by the quadratic coupling, we analyze the phase distribution  of the system (\ref{mas_eq}). The quantum mechanical phase distribution of single oscillator mode is given by \cite{barnett}
\begin{equation}
P(\phi_i)=\frac{1}{2\pi}\langle\phi_i|\rho|\phi_i\rangle=\frac{1}{2\pi}\sum_{n_1,n_2=0}^{\infty}\langle n_1|\rho|n_2\rangle e^{i(n_2-n_1)\phi_i}, \label{single_phase}
\end{equation}
where $|\phi_i\rangle=\frac{1}{2\pi}\sum_{n=0}^{\infty}e^{in\phi_1}|n\rangle$ is the phase state of the $i^{th}$ oscillator. Due to the nonlinear coupling (\ref{int_ham}) with strength $J$, the composite system (\ref{mas_eq}) develops a phase preference of $\Phi=\phi_2-2\phi_1$, thus the corresponding relative phase distribution of the coupled system (\ref{mas_eq}) can be written as \cite{hush, tilley}
{\small\begin{align}
P(\Phi)&=\frac{1}{2\pi}\sum_{n_1,n_2=0}^{\infty}\sum_{k=max(n_1,n_2)}^{\infty}e^{i(n_2-n_1)\Phi}\langle 2n_1, k-n_1|\rho|2n_2, k-n_2\rangle , \nonumber\\
&=\frac{1}{2\pi}+\frac{1}{\pi}\text{Re}\left[\sum_{p=1}^{\infty}e^{ip\Phi}\sum_{n_1=0, n_2=0}^{\infty}\rho_{n_1,n_2}^{(p)}\right], \label{phase_relative}
\end{align}}
where $\rho_{n_1,n_2}^{(p)}=\langle n_1,n_2+p|\rho|n_1+2p,n_2\rangle$ collects the off-diagonal elements associated with $p^{th}$ order $2:1$ phase correlations. 

\subsection{Perturbation analysis and Phase-locking dynamics}
To obtain the steady states of the system we utilize the perturbation analysis by writing the master equation (\ref{mas_eq}) in the number state basis $\rho_{n_1,n_2}^{(p)}=\langle n_1,n_2+p|\rho|n_1+2p,n_2\rangle$, treating the interaction term in Eq. (\ref{int_ham}) as a perturbation. From this, we can obtain the $p^{th}$ order correction to steady state density matrix as 
\begin{subequations}
\begin{align}
\rho_{n_1, n_2}^{(p)}& =\frac{1}{\lambda_{n_1, n_2}^{(p)}} \left( \mu_{1}^{(p)} + \mu_{2}^{(p)}\right),\\
\text{where}\nonumber\\
\mu_{1}^{(p)}&=-i\Omega \left( \sqrt{n_2 + p} \, \rho_{n_1, n_2}^{(p-1)} - \sqrt{n_2 + 1} \, \rho_{n_1, n_2 + 1}^{(p-1)} \right),\nonumber\\ 
\mu_{2}^{(p)}&=-iJ \left( \sqrt{(n_1 + 1)(n_1 + 2)(n_2 + p)} \, \rho_{n_1 + 2, n_2}^{(p-1)} \right.\nonumber\\
 &~~~~\left.- \sqrt{(n_1 + 2p)(n_1 + 2p -1)(n_2 + 1)} \, \rho_{n_1 , n_2 + 1}^{(p-1)} \right),\nonumber\\
\lambda_{n_1,n_2}^{(p)}&=-ip\delta - \Gamma_{n_1,n_2},
 \label{p_stesta}
\end{align}
\end{subequations}
with $\Gamma_{n_1,n_2}=\frac{\gamma_1}{2}[2(n_1+n_2)+3p+4]+\frac{\gamma_2}{2}[n_1(n_1-1)+(n_1+2p)(n_1+2p-1)+n_2(n_2-1)+(n_2+p)(n_2+p-1)]$ being the effective damping term and  $\delta=\Delta_2-2\Delta_1=\omega_2-2\omega_1$. Also $\mu_1$ and $\mu_2$ are arising from the interaction terms. In the absence of the interaction $J=0$, the steady state density matrix can be written as $\rho^{(0)}=\rho_1^{(0)}\otimes \rho_2^{(0)}$, where in the limit $\gamma_2/\gamma_1\to\infty$ each mode predominantly occupies the lowest Fock state with $\rho_i^{(0)}=(2/3)|0\rangle\langle 0|+(1/3)|1\rangle\langle 1|$ \cite{dodo}. Using  $\rho^{(0)}$ as the unperturbed state, the first order contribution from the interaction terms to density matrix read
\begin{eqnarray}
\mu_{1}^{(1)}&=&-i\Omega \, \sqrt{n_2 + 1} \left(  \rho_{n_1, n_2}^{(0)} -  \rho_{n_1, n_2 + 1}^{(0)} \right)\nonumber\\ 
\mu_{2}^{(1)}&=&-iJ \, \sqrt{(n_1 + 1)(n_1 + 2)(n_2 + 1)}\left(  \, \rho_{n_1 + 2, n_2}^{(0)} -  \, \rho_{n_1 , n_2 + 1}^{(0)} \right).\nonumber\\ \label{1_stesta}
\end{eqnarray}

This implies that, there is a nonzero contribution of the first order components to the density matrix. Substituting the non-zero first order components $\rho_{n_1,n_2}^{(1)}$ into the relative phase distribution in Eq. (\ref{phase_relative}) we obtain
\begin{equation}
P(\Phi)=\frac{1}{2\pi}\left(1+\epsilon_1\Omega+\epsilon_2J\right)\sin\Phi, \label{ana_phas}
\end{equation}
where $\epsilon_1$ and $\epsilon_2$ are the coefficients depending on the parameters $\gamma_1$ and $\gamma_2$. The $\sin\Phi$ dependence demonstrates that the coupled system develops a preferred relative phase $\Phi=\phi_2-2\phi_1$, characteristic of $2:1$ mutual-locking. 

The phase distribution of single oscillator modes can be obtained by taking the partial trace of the density matrix $\rho_{n_1,n_2}^{(1)}$ over the other mode. For first and second oscillator modes the first order correction to density matrix takes the form
{\small\begin{align}
 \rho_1^{(1)}&=Tr_{a_2}[\rho_{n_1,n_2}^{(1)}] \nonumber\\
 &= -iJ\sqrt{(n_1+1)(n_1+2)}
 \left(\sum_{n_2}\frac{\sqrt{n_2+1}}{\lambda_{n_1,n_2}}\right)
 \left[\rho_{1,n_1+2,n_1+2}^{(0)}-\rho_{1,n_1,n_1}^{(0)}\right],\nonumber\\
 \rho_2^{(1)}&=Tr_{a_1}[\rho_{n_1,n_2}^{(1)}]\nonumber\\
 &=-i\Omega\sqrt{n_2+1}\left(\sum_{n_1}\frac{1}{\lambda_{n_1,n_2}}\right)\left[\rho_{2,n_2,n_2}^{(0)}-\rho_{2,n_2+1,n_2+1}^{(0)}\right]\nonumber\\ &~~~-iJ\sqrt{n_2+1}\left(\sum_{n_1}\frac{\sqrt{(n_1+1)(n_1+2)}}{\lambda_{n_1,n_2}}\right)
 \left[\rho_{2,n_2,n_2}^{(0)}-\rho_{2,n_2+1,n_2+1}^{(0)}\right] \label{mode_dm}.
 \end{align}}
The phase distribution of single oscillator modes can be obtained by substituting $\rho_{i,n_1,n_2}$ in Eq. (\ref{single_phase}). We note that, in Eq. (\ref{mode_dm}) the first order components $\rho_{n_1,n_2}^{(1)}$, with $(n_2-n_1)=\pm 1$ from interaction terms contribute to the  perturbative steady state of first oscillator, $\rho_1^{(1)}=Tr_{a_2}[\rho_{n_1,n_2}^{(1)}]$, connects $\Delta n_1=\pm 2$ state.  And the perturbative state of the second oscillator mode connects state with $\Delta n_2=\pm 1$. These respective coherences generate the phase distributions in the form
\begin{eqnarray}
P(\phi_1)&=&(\eta_0\Omega^2+\eta_1\Omega J+\eta_2J^2)\cos 2\phi_1,\nonumber\\
P(\phi_2)&=&(\zeta_1\Omega+\zeta_2J)\sin \phi_2. \label{phase_sing}
\end{eqnarray}
where $\eta_0, \eta_1, \eta_2, \zeta_1$ and $\zeta_2$ are the constants determined by $\gamma_1$ and $\gamma_2$. From Eq. (\ref{phase_sing}) we can infer two crucial points: first, the undriven oscillator (first oscillator) develops a bistable $2:1$ phase preference and second, the driven oscillator (second oscillator) exhibits a monostable $1:1$ phase preference relative to the drive.

These analytical forms derived in Eqs. (\ref{ana_phas}) and (\ref{phase_sing}) are confirmed by the numerical phase distributions demonstrated in Fig. \ref{phase_dist} as a function of the coupling strength  in the low excitation subspace $(\Omega<\gamma_1)$ (steady-states are obtained from QuTip solver \cite{johan1, johan2}). From Fig \ref{phase_dist}(a) and \ref{phase_dist}(b), we can observe that the phase distribution of the first oscillator mode displays two maximas at $\phi_1=\pi/2$ and $3\pi/2$, reflecting $2:1$ locking and bistability  whereas the second oscillator phase-distribution shows a single maximum at $\phi_2=\pi/2$ (and minimum at $\phi_2=3\pi/2$), consistent with $1:1$ locking at the second harmonic resonance condition $\Delta_2=\omega_2-2\omega_d=0$. The relative phase distribution of the system (\ref{mas_eq}) has a single maximum at $\Phi=\pi/2$ as implied in Eq. (\ref{ana_phas}) corresponding to a monostable $2:1$ phase locking mechanism between the oscillators as evidenced in Eq. (\ref{ana_phas}). The classical mean-field analysis presented in Appendix A reproduces this asymmetric phase-locking geometry, yielding two symmetry-related $2:1$ phase-locked states for the first oscillator ($\phi_1=\pi/2$ and $3\pi/2$) and a $1:1$ phase-locked state for the second oscillator ($\phi_2=3\pi/2$). However, while the classical equations describe the phase-locking structure, they do not explain the microscopic origin of the absence of conventional 
$1:1$ synchronization of the first oscillator. This origin is revealed by the perturbative quantum analysis, where the quadratic interaction imposes a two-phonon selection rule that suppresses direct single-phonon excitation of the first oscillator, thereby blocking the conventional $1:1$ synchronization channel while allowing higher-order $2:1$ synchronization. We will discuss this in Sec. \ref{synch_mea}.

Altogether, the perturbative analysis reveals  that the first oscillator synchronizes through the higher order $2:1$ phase relation with the drive. Whereas, the second oscillator follows conventional phase-locking pattern ($1:1$) to the drive, but only after its frequency is renormalized to $\Delta_2=\omega_2-2\omega_d$ by the quadratic interaction. This collectively gives us a high-order mutual phase-locking behavior at the resonance condition $\Delta_2=2\Delta_1$. In the following, we quantify these phase-locking behaviors using the synchronization correlators and discuss them from the perspective of underlying phonon population dynamics and dressed state structures.

\section{Synchronization measures and Phonon-statistical properties}\label{synch_mea}
\begin{figure}
	\centering\includegraphics[width=1.0\linewidth]{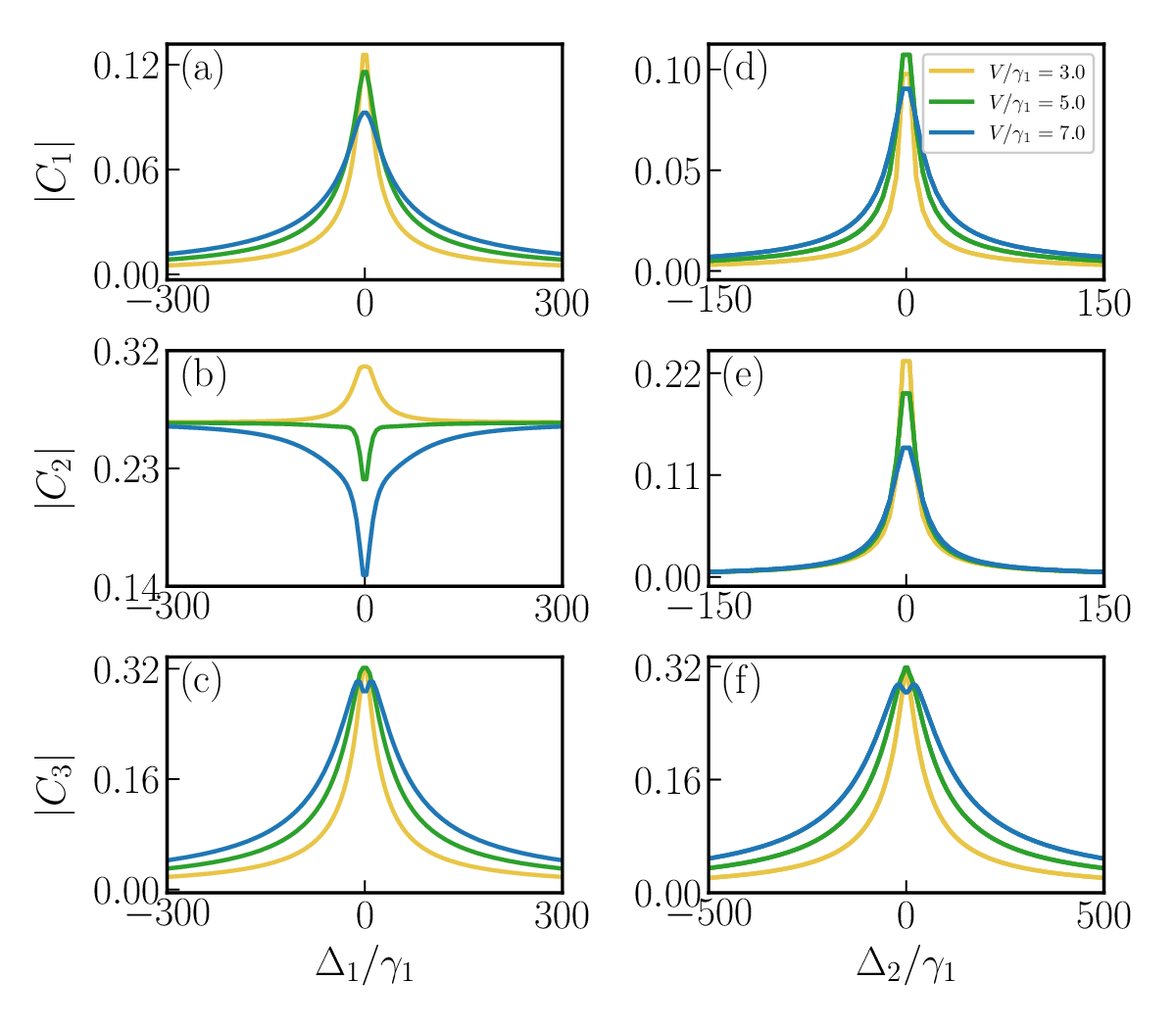}
	\caption{Absolute value of synchronization measures $C_i$ [see Eq.(\ref{synch_meas})] plotted as a function of $\Delta_1$ in panels (a)-(c) with $\Delta_2=0$ and as a function of $\Delta_2$ in panels (d)-(f) with $\Delta_1=0$ for different coupling strengths. In all these cases we have considered $\Omega/\gamma_1=0.5$ and $\gamma_2/\gamma_1=10$.}
	\label{synch2d}
\end{figure}
\begin{figure*}
	\includegraphics[width=0.85\linewidth]{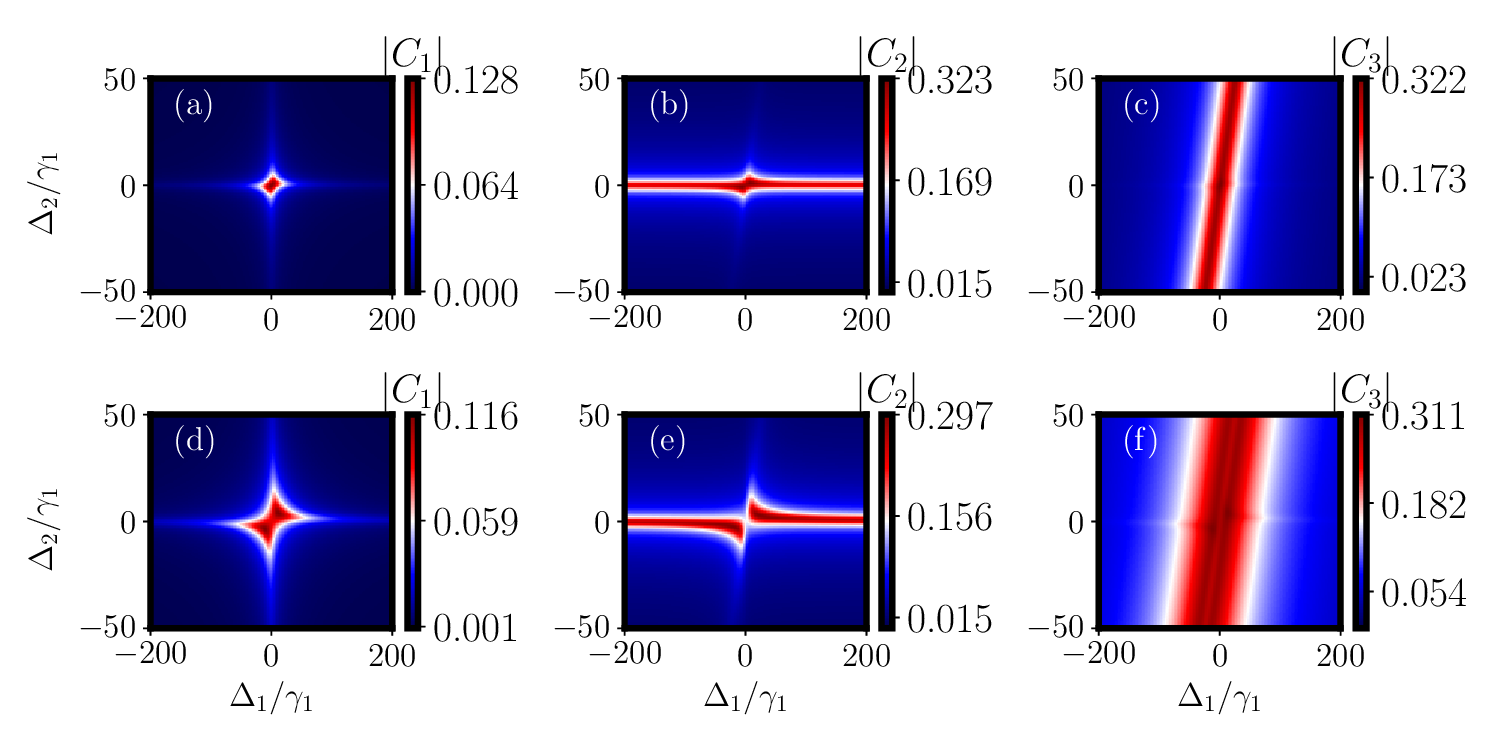}
	\caption{Phase locking measures $|C_i|$ defined in Eq. (\ref{synch_meas}) of system (\ref{mas_eq}) plotted as a function of $\Delta_1$ and $\Delta_2$ for $J/\gamma_1=3.0$ in panels (a)-(c) and for $J/\gamma_1=7.0$ in panels (d)-(f) with $\Omega/\gamma_1=0.5$ and $\gamma_2/\gamma_1=10$. }
	\label{surface_rel}
\end{figure*}
To quantify synchronization measure we adopt the synchronization correlator which compresses the information in phase distribution into a single quantity. As discussed in the previous section, the first oscillator undergoes synchronization blockade with the drive and from Eq. (\ref{phase_sing}) we find that the phase relation between the oscillator and the drive is of the order $2:1$ and the second oscillator undergoes $1:1$ phase-locking with the drive. Also, the oscillator mutually phase-lock with each other in the order $2:1$, as obtained in Eq. (\ref{phase_relative}). Based on this, we can define the synchronization measures as \cite{weiss, nissi}
\begin{eqnarray}
C_1&=&|C_i|e^{-2\phi_1}=\frac{\langle a_1^{\dagger 2}\rangle}{\sqrt{\langle a_1^{\dagger}a_1\rangle}}, \nonumber\\
C_2&=&|C_i|e^{-\phi_2}=\frac{\langle a_2^{\dagger}\rangle}{\sqrt{\langle a_2^{\dagger}a_2\rangle}}, \nonumber\\
C_3&=&|C_3|e^{-i\Phi}=\frac{\langle a_1^{\dagger 2}a_2\rangle}{\sqrt{\langle a_1^{\dagger}a_1\rangle\langle a_2^{\dagger}a_2\rangle}}. \label{synch_meas}
\end{eqnarray} 

Here we characterize the synchronization using the absolute value of the measures $|C_i|$. Figure \ref{synch2d} demonstrates the absolute value of synchronization measures $C_i$ plotted as a function of $\Delta_{1,2}$. Figures \ref{synch2d}(a) and \ref{synch2d}(d) demonstrates the absence of conventional $1:1$ synchronization of undriven oscillator, As predicted by Eq. (\ref{phase_sing}), the first oscillator does not show $1:1$ phase-locking with the external drive not only at the resonance but across entire parameter space of $\Delta_{1,2}$ and for every coupling strength $J$. Instead it displays high-order $2:1$ phase-locking dynamics that constitutes the primary signature of blockade induced by the quadratic coupling. 
Specifically in Fig. \ref{synch2d}(a), with respect to $\Delta_1$ with $\Delta_2$ fixed at zero, the resonant peaks of the synchronization measure $|C_1|$ that quantify $2:1$ locking shows decreasing magnitude with increasing coupling strength $J$ . In contrast, Fig. \ref{synch2d}(d), in response to the detuning $\Delta_2$ with fixed $\Delta_1$, shows the resonant peaks of the synchronization measure $|C_1|$, that initially increases with the coupling strength, reaches optimal $2:1$ coherence at intermediate $J$ before subsequently decreasing at very strong coupling.  Figure \ref{synch2d}(b) and \ref{synch2d}(e) shows synchronization measure of second oscillator with the drive $|C_2|$ against $\Delta_1$ when $\Delta_2=0$ and against $\Delta_2$ when $\Delta_1=0$ respectively. The synchronization measure $|C_2|$ represents $1:1$ phase-locking with the drive. In Fig. \ref{synch2d}(b) we can observe that for lower coupling strengths, $C_2$ shows resonant synchronization peaks centered at $\Delta_1=0$. However, as the coupling strength $J$ is increased we can see that resonant peaks transforms into synchronization dips at the resonance ($\Delta_1=0$), while $|C_2|$ maintains nearly constant non-zero value across the off-resonant regime demonstrating that the driven oscillator continues to exhibit $1:1$ phase coherence but at the frequency $\omega_2=2\omega_d$ rather than the expected frequency $\omega_2=\omega_d$. Whereas, the dynamics of $|C_2|$ against $\Delta_2$ as shown in Fig. \ref{synch2d}(e) with $\Delta_1=0$, exhibits resonant peaks centered at $\Delta_2=0$ for all coupling strengths, confirming that the resonance condition for the driven oscillator corresonds to $\omega_2=2\omega_d$, although the amplitude of the resonant peaks decreases with the coupling strength $J$. Also, the relative synchronization measure $|C_3|$, which characterizes the high-order phase-relation $(2:1)$ between the oscillators emerging due to the coupling,  shows a similar phase-locking behavior against $\Delta_1$ and $\Delta_2$ as illustrated in Figs. \ref{synch2d}(c) and \ref{synch2d}(f) respectively. The magnitude of the peaks increases with the coupling strength. As the coupling strength is increased further, a slight dip in the resonance starts to emerge along with the off-resonant peaks in both the cases as depicted in the Figs. \ref{synch2d}(c) and \ref{synch2d}(f). 

\subsection{Phonon statistics}
\begin{figure*}
	\includegraphics[width=0.85\linewidth]{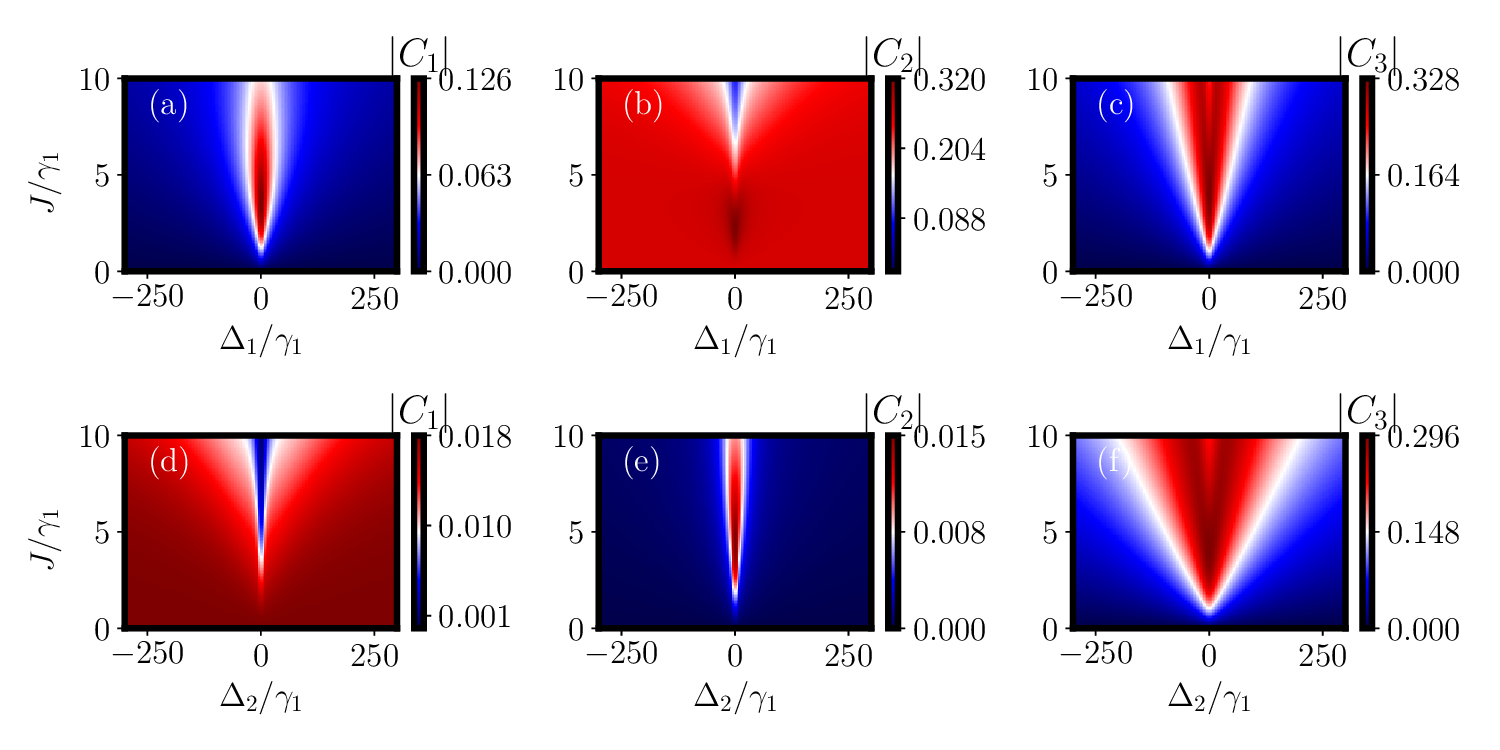}
	\caption{Regimes of synchronization of system (\ref{mas_eq}). The absolute value of synchronization measures $C_i$ defined in Eq. (\ref{synch_meas}) plotted as a function of coupling strength $J/\gamma_1$ and detuning $\Delta_1/\gamma_1$ with $\Delta_2=0$ in panels (a)-(c) and detuning $\Delta_2/\gamma_1$ with $\Delta_1=0$ in panels (d)-(f). We have considered the drive strength $\Omega/\gamma_1=0.5$ and damping rate $\gamma_2/\gamma_1=10$.  } 
	\label{Arnold}
\end{figure*}
To have further insight into the synchronization mechanism observed in $C_i$, we investigate the steady-state dynamics through probability amplitudes from Schr\"{o}dinger equation approach by considering few-phonon levels in the weak excitation regime ($E\ll\gamma_1,\gamma_2$) as illustrated by energy level structure in Fig. \ref{energy}. Within the low excitation subspace, we can write the wavefunction ansatz as
\begin{equation}
|\psi\rangle=c_{00}|00\rangle+c_{10}|10\rangle+c_{20}|20\rangle+c_{01}|01\rangle, \label{ansatz}
\end{equation}
where $c_{n_1n_2}$ represents the probability amplitudes and $|c_{n_1n_2}|^2$ gives the occupational probability of finding the system in the state $|n_1n_2\rangle$. To obtain the probability amplitudes we consider a non-Hermitian Hamiltonian which incorporates the dissipative effects, $H'=\sum_{i=1}^2\Delta_ia_i^{\dagger}a_i-\frac{\gamma_1}{2}a_ia_i^{\dagger}-i\frac{\gamma_2}{2}a_i^{\dagger 2}a_i+H_I$. Using the Schr\"{o}dinger equation $i\frac{\partial |\psi\rangle}{\partial t}=H'|\psi\rangle$, the equation of motion for probability amplitudes $c_{n_1n_2}$ are obtained as
\begin{eqnarray}
i\dot{c}_{00}&=&-i\gamma_1c_{00}+\Omega c_{01},\nonumber\\
i\dot{c}_{10}&=&\left(\Delta_1-i\frac{3\gamma_1}{2}\right)c_{10},\nonumber\\
i\dot {c}_{20}&=&\left(2\Delta_1-i(2\gamma_1+\gamma_2)\right)c_{20}+\sqrt{2}Jc_{01},\nonumber\\
i\dot{ c}_{01}&=&\left(\Delta_2-i\frac{3\gamma_1}{2}\right)c_{01}+\Omega c_{00}+\sqrt{2}Jc_{20}. \label{eom}
\end{eqnarray}

In the weak excitation regime, we can obtain the steady solution from Eq. (\ref{eom}) with  $c_{00}\approx1$ and find that the steady-state coefficient $c_{10}$ in the first excitation regime is a constant. The steady-state probability amplitudes in the second excitation regime are approximately
\begin{eqnarray}
c_{20}=\frac{\sqrt{2}J\Omega}{\left(\Delta_2-i\frac{3\gamma_1}{2}\right)\left(2\Delta_1-i(2\gamma_1+\gamma_2)-2J^2\right)},\nonumber\\
c_{01}=\frac{-\Omega\left(2\Delta_1-i(2\gamma_1+\gamma_2)\right)}{\left(\Delta_2-i\frac{3\gamma_1}{2}\right)\left(2\Delta_1-i(2\gamma_1+\gamma_2)-2J^2\right)}. \label{sspa}
\end{eqnarray}
\begin{figure*}
	\includegraphics[width=0.85\linewidth]{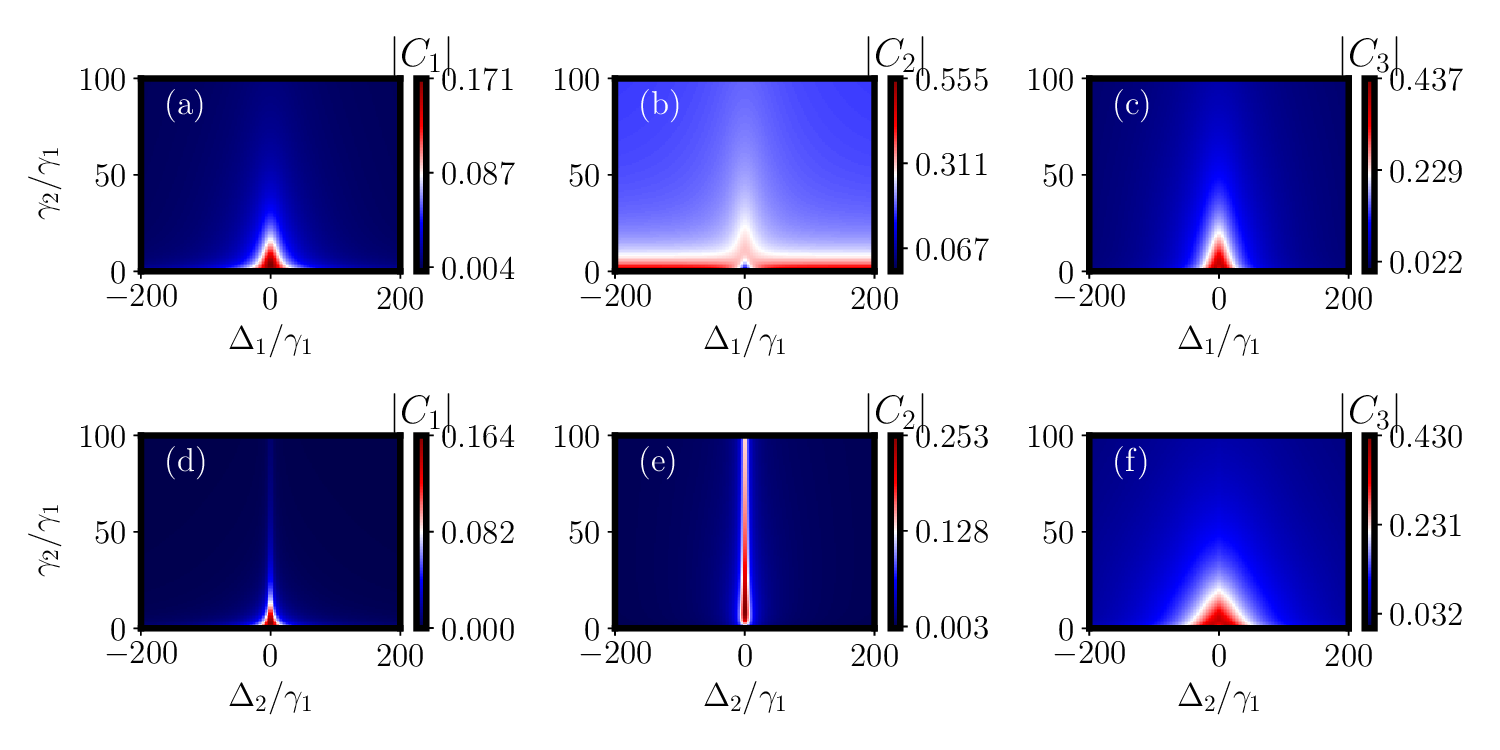}
	\caption{Absolute value of synchronization measures $C_i$ [Eq. \ref{synch_meas}] as a function of damping rate $\gamma_2/\gamma_1$ and detunings $\Delta_1/\gamma_1$ in panels (a)-(c) and $\Delta_2/\gamma_1$ in panels (d)-(f) for coupling strength $J/\gamma_1=3.0$ and drive strength $\Omega/\gamma_1=0.5$.   }
	\label{damping1}
\end{figure*}

These analytical expressions provide the basic mechanism of the synchronization dynamics of the system (\ref{mas_eq}). The classical steady-state analysis presented in Appendix A reproduces the asymmetric phase-locking behavior of the system, namely the two phase-locked states of the first oscillator and the $1:1$ phase-locked state of the second oscillator. While the classical analysis establishes the phase-locking configuration of the system, the perturbative Schr\"{o}dinger analysis presented here provides the microscopic picture underlying the synchronization dynamics. Since the first oscillator is neither directly driven nor coupled to the single-phonon manifold through the quadratic interaction, the first excited state $|10\rangle$ remains unpopulated in the weak exciation regime as evident from then second equation of Eq. (\ref{eom}). Consequently, the conventional first-harmonic synchronization is suppressed, whereas synchronization proceeds exclusively through the two-phonon state $|20\rangle$, giving rise to higher-order $2:1$ synchronization.  In contrast, the second oscillator experiences direct driving that efficiently populates the excited state $|01\rangle$, through the direct transition  $|00\rangle \to |01\rangle$ with probability amplitude $c_{01}$ given in Eq. (\ref{sspa}) enabling $1:1$ phase-locking at the frequency $\omega_2=2\omega_1$. However, the coupling introduces a second  process - a coherent transition between the bare states $|10\rangle$ and $|20\rangle$ which happens in the second excited state when these states becomes energitically accessible, which initiates the high-order synchronization dynamics in the undriven oscillator while simultaneously revealing how the coupling affects the synchronization of first and second oscillator with the drive through the explicit relationship between the probability amplitudes $c_{20}$ and $c_{01}$. From Eq. (\ref{sspa}), we can obtain a relation
\begin{equation} c_{20}=\left(-\frac{\sqrt{2}J}{2\Delta_1-i(2\gamma_1+\gamma_2)}\right)c_{01}, \label{pop_exc}
\end{equation} 
which implies that for the condition $J\gg2\Delta_1,2\gamma_1+\gamma_2$, the phonon population transfers from second oscillator mode $|01\rangle$ to first oscillator's two phonon state $|20\rangle$. This coupling induced tunneling of the phonons in the second excited state allows the observed $2:1$ synchronization enhancement of first oscillator with the drive and suppression of $C_2$ resonant peaks. This resonant phonon tunneling in the second excitation manifold enables phase-locking of the first oscillator with the drive, matching the $2:1$ synchronization peaks observed in the Figs. \ref{synch2d}(a) and \ref{synch2d}(d). When the frequency of the second oscillator ($\omega_2$) becomes twice the frequency of the drive ($2\omega_d$) the direct drive transition of the phonons from ground state $|00\rangle$ to second excited state $|01\rangle$ becomes maximally efficient, enabling  $1:1$ phase-locking between the second oscillator and the drive as illustrated in Figs. \ref{synch2d}(b) and \ref{synch2d}(e) in the weak coupling regime where the phonon population remains predominantly in the second oscillator mode. However, as the coupling strength is increased beyond $J> \gamma_1$ with $\Delta_1\approx0$,  the phonons start to tunnel to the undriven mode $|20\rangle$ causing characteristic synchronization dips to emerge at the resonance. In the off resonant regime with $|2\Delta_1|\gg J$, we can observe the constant synchronized response across broad detuning ranges, since the phonon transfer between the modes becomes weaker. Furthermore, in response to the direct detuning $\Delta_2$, the  measure $|C_2|$ shows strong resonant synchronzation peaks and the magnitude of these peaks progressively decreases with the coupling strength as a result of the phonon transfer process, as shown in Fig. \ref{synch2d}(e). 

The mutual synchronization between the oscillators occurs when resonance condition, $\delta=\Delta_2-2\Delta_1=\omega_2-2\omega_1=0$, is satisfied enabled by the transition between the ground and second excited state and also between the bare $|20\rangle\to|01\rangle$ states. When the condition, $J>\gamma_1,\gamma_2$ is reached at the resonance  $\Delta_1,\Delta_2=0$, the bare eigenstates $|20\rangle$ and $|01\rangle$ split into two distinct nondegenerate dressed states $|\Psi^{\pm}\rangle$ whose energy separation induce slight suppression of the mutual synchronization measure $|C_3|$ at the resonance $\delta=0$, which becomes prominent with increasing coupling strength along with second synchronization peaks at $\delta=\pm2\sqrt{2}J$ as clearly depicted in Figs. \ref{synch2d}(c) and \ref{synch2d}(f). We have discussed the synchronization mechanism of system (\ref{mas_eq}) without the drive in \cite{nissi}.

\subsection{Regimes of synchronization}
\begin{figure*}
	\includegraphics[width=0.85\linewidth]{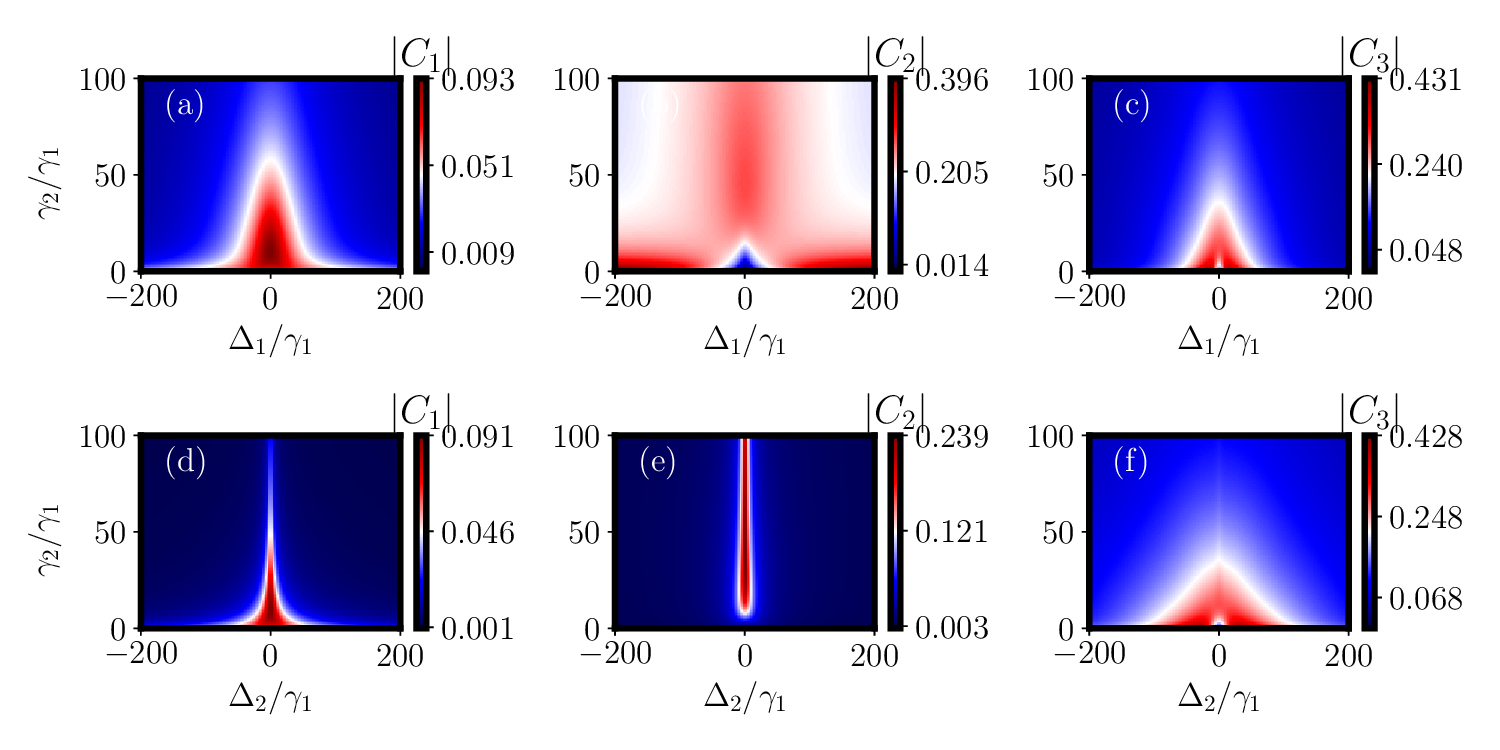}
	\caption{Absolute value of synchronization measures $C_i$, see Eq. (\ref{synch_meas}),  as a function of damping rate $\gamma_2/\gamma_1$ and detunings $\Delta_1/\gamma_1$ in panels (a)-(c) and $\Delta_2/\gamma_1$ in panels (d)-(f) for coupling strength $J/\gamma_1=7.0$ and drive strength $\Omega/\gamma_1=0.5$. }
	\label{damping2}
\end{figure*}
 To present a complete dynamical picture, we illustrate the synchronization measures as a function of system parameters. Figure \ref{surface_rel}, shows the response of synchronization measures $|C_i|$ in Eq. (\ref{synch_meas}) as the function of detunings $\Delta_1$ and $\Delta_2$. In weak coupling regime ($J=3.0\gamma_1$), we can observe phase-locking region centered around $\Delta_1=\Delta_2=0$ in Fig. \ref{surface_rel}(a), demonstrating that the undriven oscillator achieves $2:1$ synchronization through the limited phonon tunneling that weakly couples the driven $|01\rangle$ state to the undriven state $|20\rangle$ while satisfying the required resonance condition. Whereas at the strong coupling regime $J=7.0\gamma_1$, Fig. \ref{surface_rel}(d) shows expanded synchronization regime across both $\Delta_1$ and $\Delta_2$ because  enhanced coupling significantly increases phonon population transfer from driven to undriven oscillator modes, enabling $2:1$ phase-locking over broader range of detunings.
 
 For the coupling strength $J/\gamma_1=3.0$, we can observe a horizontal bright band around $\Delta_2=0$ extending across all $\Delta_1$ range as shown in Fig. \ref{surface_rel}(b) demonstrating that the driven oscillator achieves strong $1:1$ synchronization with the external drive at the resonace condtion $\Delta_2=0$ regardless of the detuning $\Delta_1$. This is because in the weak coupling regime, the phonon tunneling remains weak and the phonon population remains in the driven mode. However, at the strong coupling regime ($J=7.0\gamma_1$) Fig. \ref{surface_rel}(e) shows the horizontal synchronization band which develops a pinched or split structure at $\Delta_1=0$ with a vertical line appearing at the intersection of the horizontal $\Delta_2=0$ band. This vertical line appears at the resonance point $(\Delta_1, \Delta_2)=(0,0)$ where maximum phonon tunneling between the degenerate states $|20\rangle$ and $|01\rangle$ depletes the driven oscillator population while surrounding bright band confirms the $1:1$ locking at $\omega_2=2\omega_d$. 

 Furthermore, in this regime, the mutual synchronization between the oscillators shows a diagonal bright band which is demonstrated in Fig. \ref{surface_rel}(c) and this signifies that the oscillators are synchronized strongly when the frequency mismatch ($\delta=\Delta_2-2\Delta_1=\omega_2-2\omega_1$) between the oscillators are near resonance.  This indicates that the coupling has reached the condition ($J>\Delta_1, \gamma_1$) when the population transfer between the modes $|01\rangle$ and $|20\rangle$ is more effective and the probability of finding the phonons in driven oscillator mode is less than the undriven mode at the resonance $\Delta_1=0$. Besides this, the mutual synchronization regime becomes broader in the strong coupling regime as displayed in Fig. \ref{surface_rel}(f) as the maximum synchronization is along $\delta=\pm2\sqrt{2}J$, because the degeneracy of the states $|20\rangle$ and $|01\rangle$ is lifted and synchronization is slightly supressed along $\delta=0$. 

Figure \ref{Arnold} shows the synchronization regimes of the  complete system (\ref{mas_eq}). Here we have illustrated the response of phase-locking measures ($|C_i|$) to the coupling strength $J$ and the detunings $\Delta_{1,2}$. In Fig. \ref{Arnold}(a) we can observe the phase-locking regime of the first oscillator with respect to the drive, represented by the measure $|C_1|$, which increases with the coupling strength because of the effective phonon transfer to the first mode and the magnitude gradually decreases in the strong coupling regime when the bare states start to split into dressed state. With respect to the detuning $\Delta_2$, the first oscillator shows synchronization regime in the lower values of the coupling and decreases with the coupling as shown in Fig. \ref{Arnold}(d). Figure \ref{Arnold}(b) shows the phase-locking regime of second oscillator with the drive where, for lower values of coupling strength there is a small regime around the resonance $\Delta_1=0$ where the second oscillator is strongly phase locked with the drive and outside that regime we can observe constant synchronized behavior. As the coupling strength is increased, there is a transition from strong synchronized behavior to weak sychronization regime which broadens with the coupling. That is when the coupling strength crosses the condition $J>2\Delta_1, 2\gamma_1$, the population drop in the mode $|01\rangle$ weakens the synchronization of the second oscillator with the drive. So, instead of resonant peaks we observe dips near zero detuning which reduces the driven oscillators direct synchronization with the drive. With its own detuning $\Delta_2$, the synchronization of the second oscillator with the drive is maximal near the resonance as shown in Fig. \ref{Arnold}(c) and it broadens with the coupling, but it's magnitude drops in the strong coupling regime. In Figs. \ref{Arnold}(c) and \ref{Arnold}(f) we can observe ``Arnold-tongue" shaped synchronized regimes which broadens with the coupling strength. We can also observe the split at the resonance where maxima of synchronization shifts from the resonace $2\Delta_1=\Delta_2=0$ to $\delta=\pm2\sqrt{2}J$ in Figs. \ref{Arnold}(c) and \ref{Arnold}(f).

The dissipation rates affects population transfer as we have seen from Eq. (\ref{sspa}) and we have demontrated its impact in Figs. \ref{damping1} ($J/\gamma_1=3.0$) and \ref{damping2} ($J/\gamma_1=7.0$). In lower coupling regime, when the dissipation rate is low, with $\gamma_2/\gamma_1<J$, we can observe the phase-locking region which disappear for higher dissipation rates which is demonstrated in Fig. \ref{damping1}(a) and \ref{damping1}(d). We illustrate the regime of synchronization dips near resonance in Fig. \ref{damping1}(b), which alters to synchronization regime when $\gamma_2/\gamma_1>J$. This resonant spike gets suppressed for higher dissipation rates. Also for lower dissipation, synchronization of the second oscillator is strongest around resonance ($\Delta_2=0$), whose magnitude decreases with increased damping rates as shown in Fig. \ref{damping1}(e). In Figs. \ref{damping1}(c) and \ref{damping1} (f) the mutual synchronization is strongest for lower damping rates, and diminishes for higher values. In the stronger coupling regime, the synchronization regime persists for higher values of dissipation rates as displayed in Fig. \ref{damping2} and the synchronization regimes are also broader than the lower coupling regimes. Even for the synchronization for the second oscillator with the drive to persist for higher dissipation around the resonance a stronger coupling strength is necessary to overcome the dissipation as demonstrated in Fig. \ref{damping2}(e). Furthermore, higher damping suppresses mutual synchronization as shown in Figs. \ref{damping2}(c) and \ref{damping2}(f) but the effect is less severe than the lower coupling. This shows that, sufficient quadratic coupling strength can maintain the phase-locking at second harmonic frequency against strong dissipation rates. Therefore, for strong coupling rates synchronization is more robust to detuning and damping.
 
\section{Power spectrum Analysis}\label{power_spe}
\begin{figure*}
	\includegraphics[width=0.75\linewidth]{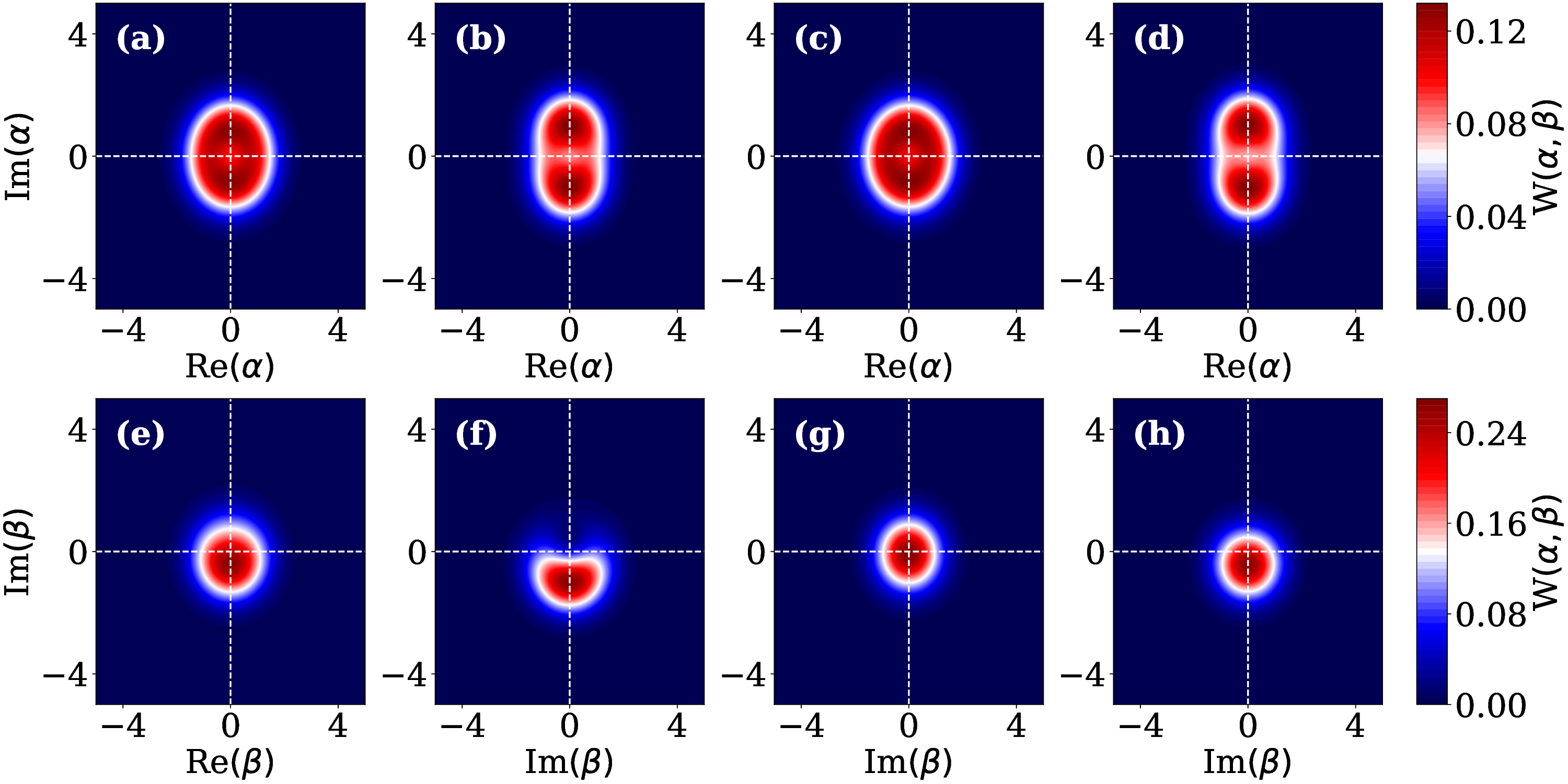}
	\caption{Wigner function of steady-state of oscillator modes. Panels (a)-(d) represents the Wigner function of the first oscillator and panels (e)-(h) represents the Wigner function of second oscillator mode. We have plotted the limit cycle for coupling strength $J/\gamma_1=3.0$ with $\Omega/\gamma_1=0.5$ in panels (a) and (e) and $\Omega/\gamma_1=3.0$ in (b) and (f). Limit cycles for coupling strength $J/\gamma_1=7.0$ is plotted with $\Omega/\gamma_1=0.5$ in panels (c) and (g) and $\Omega/\gamma_1=3.0$ in (d) and (h). In all these figures we have considered, $\Delta_1=\Delta_2=0$ and $\gamma_2/\gamma_1=10$. }
	\label{wigner}
\end{figure*}
So far we have analyzed the dynamics in the low excitation regime by restricting to weak driving such that the system remains predominantly in lowest phonon manifolds. In this section we visualize the behavior of each oscillators in phase space using the Wigner function and analyze the frequency entrainment using Power spectrum for both low and high drive strengths to obtain a more clearer picture of the underlying dynamics. 

In the weak coupling regime, the undriven oscillator acquires only a small but finite population transfer from the driven mode induced by the coupling in the low excitation manifold. As a result we can observe a oval shaped Wigner function in Fig. \ref{wigner}(a). In the same regime, Fig. \ref{wigner}(e) represents the Wigner distribution of the driven oscillator (second oscillator) which acquires the phonon population directly from the drive. Here we observe that the Wigner distribution is offset from the origin as the drive acts as a coherent non-zero displacement in the phase-space. When the drive strength is increased at the same coupling regime the Wigner function develop a pinched distribution as shown Fig. \ref{wigner}(b) while the second (driven) oscillator loses its radial symmetry as illustrated in Fig. \ref{wigner}(f).  The pinching arises from quadrature squeezing induced by an effective quadratic drive on the first oscillator. Under strong driving, the second oscillator undergoes phase pulling and develops a large coherent amplitude, $\langle a_2 \rangle = \alpha$, as evidence by the offset Wigner function in Fig. \ref{wigner}(f). Substituting this mean-field amplitude $a_2 = \alpha + \delta a_2$ into the interaction Hamiltonian (\ref{int_ham}) yields effective Hamiltonian for the first oscillator, \cite{walls}
\begin{equation}
H_{\text{eff}}^{(1)} \simeq J(\alpha a_1^{\dagger 2} + \alpha^* a_1^2). \label{eff_ham}
\end{equation}
This Hamiltonian thus induces the quadrature squeezing in the first oscillator, appearing as pinched Wigner distribution in Fig. \ref{wigner}(b). In the strong coupling, low drive regime the first oscillator shows an oval Wigner distribution in Fig. \ref{wigner}(c) while the second oscillator maintains radial symmetry around the origin as shown in Fig. \ref{wigner}(g). Here, coupling dominates drive, maximizing population transfer to first oscillator and suppressing drive-induced displacement in the second oscillator. At high driving regime (same coupling strength), Fig. \ref{wigner}(d) shows a more refined pinched distribution from enhanced quadrature squeezing. At the same time, the Wigner function of the second oscillator becomes radially symmetric but offset from the origin as shown in Fig. \ref{wigner}(h). 

\begin{figure}
	\includegraphics[width=1.0\linewidth]{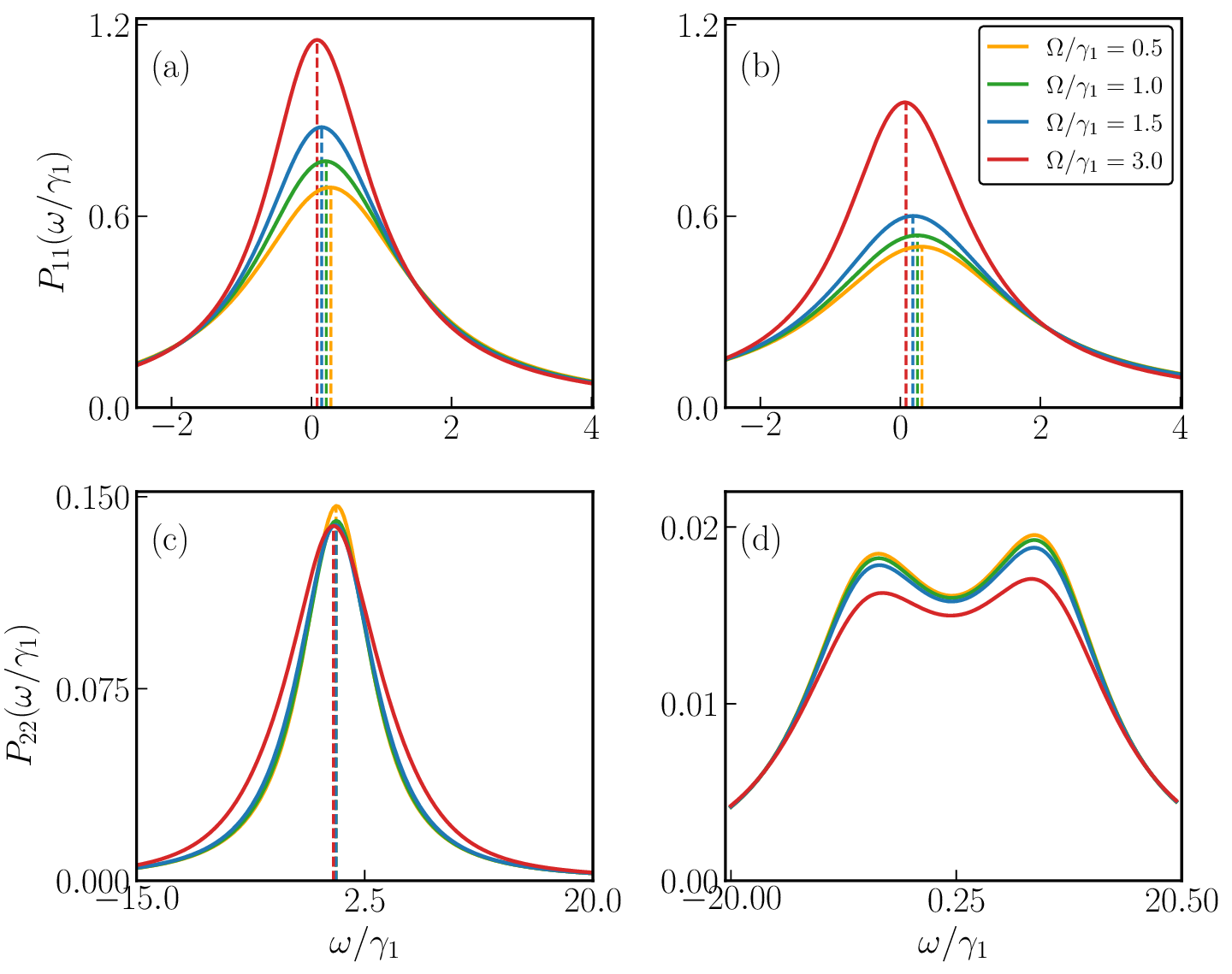}
	\caption{Power spectrum defined in Eq. (\ref{power}) of the system (\ref{mas_eq}). The spectrum of the first oscillator, $P_{11}(\omega)$ is plotted for (a)  $J=3.0$ and (b)  $J=7.0$ with $\Delta_1=0.3\gamma_1$ and $\Delta_2=0$. Power spectrum of second oscillator $P_{22}(\omega)$, plotted for (c)  $J=3.0$ and (d)  $J=7.0$ for $\Delta_1=0\gamma_1$ and $\Delta_2=0.3\gamma_1$. In all these plots we have considered $\gamma_2/\gamma_1=10$. }
	\label{spec}
\end{figure}
Consistent with this picture, we determine the role of strong driving in the frequency domain using the Power spectrum defined as 
\begin{equation}
P_{ii}(\omega)=\int_{-\infty}^{\infty}dt e^{i\omega t}\langle a_i^{\dagger}(t)a_i(0)\rangle,\qquad i=1,2. \label{power}
\end{equation}
where $P_{ii}$ represents the spectrum of the $i^{th}$ oscillator. The spectra of first oscillator $P_{11}(\omega)$ in Fig. \ref{spec}(a) reveals a sharper subharmonic peaks with increasing drive inducing strong entrainment due to quadrature squeezing in the weak coupling regime as explained above. In the strong coupling regime, from Fig. \ref{spec}(b), we can observe that the coupling slows the initial entrainment through collective hybrid modes (splitting of the bare states into dressed states), but yields sharper spectral peaks for strong drive as squeezing is enhanced. Conversely, in the low coupling regime, we observe that the spectrum of the second oscillator $P_{22}(\omega)$ exhibits suppression as the drive strength is increased as shown in Fig. \ref{spec}(c). Here the oscillator is weakly entrained. When the coupling is increased we can observe split in the spectrum characteristics of normal mode splitting between the hybridised states ($|01\rangle$ and $|20\rangle|$).
\section{Experimental realization}\label{exp_rea}
Experimental realization of Eq. (\ref{mas_eq}) can be achieved in a trapped-ion setup. In this scheme, two motional modes with frequencies $\omega_0$ and $2\omega_0$ are used through their side-band transition to implement the system (\ref{mas_eq}). By exciting the blue sideband transitions of the two modes with properly chosen detunings, one can realize the dynamics in Eq. (\ref{mas_eq}) and regulate both the linear phonon gain ($\gamma_1$) and nonlinear loss ($\gamma_2$) \cite{lee}. The interaction Hamiltonian (\ref{int_ham}) can be realized by exploiting the intrinsic nonlinearity between ions, which is activated when the trapped frequencies of different motional modes are tuned into the resonance condition ($\omega_s=2\omega_r$). Under this condition and by applying an RF field to the second mode, the interaction Hamiltonian can be generated \cite{ding17}. 

A different realization of the quantum van der Pol oscillator can be achieved in optomechanical systems operating in membrane in the middle configuration. In this architecture, quadratic optomechanical interactions enable processes involving the simultaneous absorption or emission of two phonons \cite{walter14}. When the optical field is driven on the red two-phonon sideband, the nonlinear damping ($\gamma_2$) of the mechanical motion is achieved. Conversely, driving a linearly coupled cavity mode on blue one-phonon sideband leads to effective negative damping $(\gamma_1)$ of the system. The configuration corresponding to Eq. (\ref{mas_eq}) with the interaction Hamiltonian in Eq. (\ref{int_ham}) can be engineered by considering two mechanical oscillators that individually exhibit van der Pol dynamics. These oscillators are coupled to each other through a nonlinear quadratic interaction of strength $g$ and are also connected to a shared optical cavity via radiation-pressure coupling characterized by $g_0$ \cite{xu,li}. If one of the mechanical modes is subjected to an external coherent drive with amplitude $\Omega$, the total Hamiltonian describing the system, expressed in a reference frame rotating with the laser frequency, takes the form
\begin{eqnarray}
H=\omega_c a^{\dagger}a+\sum_{i=1,2}\Delta_{m,i}b_i^{\dagger}b_i+g_0a^{\dagger}a(b_i^{\dagger}+b_i)\nonumber\\
~~+g(b_1^{\dagger2}b_2+b_1^{2}b_2^{\dagger})+\Omega(b_2^{\dagger}+b_2)
\end{eqnarray}
where $\Delta_{m,1}=\omega_{m,1}-\omega_d$ and $\Delta_{m,2}=\omega_{m,2}-2\omega_d$ represents the detunings between the mechanical mode frequencies and the driving frequency. 
\section{Conclusion}\label{conc}
In this work, we have investigated synchronization behavior of a pair of quadratically coupled quantum oscillators subjected to external driving. Using perturbation analysis, we analytically obtained the phase distribution which revealed revealed the suppression of conventional $1:1$ phase-locking in the undriven oscillator while exhibiting robust $2:1$ phase-locking with the drive. In contrast, we identified that the driven oscillator exhibits conventional 1:1 synchronization with the external drive at the second harmonic frequency ($\omega_2=2\omega_d$). We also observed that the system displays high order $2:1$ phase-locking dynamics. To understand the energy redistribution between the oscillators, we analyzed the phonon statistical properties and quantified the resulting phase correlations using the synchronization measures. These measures demonstrated how the coupling strength modulated the synchronization regime, showing the transition from Arnold tongue structures at weak coupling to pinched synchronization regions and dressed state splitting at strong coupling. The phonon statistical properties obtained from steady-state amplitudes quantitatively revealed how the transfer of phonon population from driven to undriven oscillator governs the synchronization dynamics. The result showed that when the resonance conditions are satisfied and the coupling strength becomes sufficiently strong the maximum phonon transfer occurs. This weakens the direct coherent response of the driven oscillator and simultaneously establishing the two-phonon coherence required for $2:1$ locking in the undriven oscillator. We also find that the weak dissipation in the undriven oscillator preserves essential two-phonon coherence for high-order synchronization, while higher dissipation in the driven oscillator maintains its responsiveness to external drive, enabling the mechanism across experimental parameter regime.
We also observed the quadrature squeezing in the Wigner function of the undriven oscillator in the strong driving regime. The power spectrum revealed strong entrainment due to squeezing-enhanced coherence in this regime. 

Our findings can establish quadratic coupling as quantum selection rule for synchronization order. Here the quadratic coupling suppress the low order phase locking while amplifying high-order locking and frequency conversion. Our results may also provide insight into controlling phase coherence and energy transfer. The blockade can enable engineered phase relationships in oscillator arrays, realizable in trapped ion and optomechanical setups. 

\section*{Acknowledgements}
NT thanks Science and Engineering Research Board,  Government of India, under the Grant No. CRG/2021/002428, for providing the Senior Research Fellowship. The work of MS forms part of a research project sponsored by Science and Engineering Research Board,  Government of India, under the Grant No. CRG/2021/002428.
\appendix
\section{Classical steady-state analysis}
The classical equation of motion corresponding to Eq. (\ref{mas_eq}) can be obtained within the mean field approximation by replacing the operator $ a_j$ with their expectation value $\langle a_j\rangle=\alpha_j$ and using the relation $\frac{d\langle a_j \rangle}{dt}=Tr[a_j\dot\rho]$, which yields
\begin{eqnarray}
\dot{\alpha_1}&=&\left(-i\Delta_1+\frac{\gamma_1}{2}-\gamma_2|\alpha_1|^2\right)\alpha_1-2iJ\alpha_1^{*}\alpha_2,\nonumber\\
\dot{\alpha_2}&=&\left(-i\Delta_2+\frac{\gamma_1}{2}-\gamma_2|\alpha_2|^2\right)\alpha_2-iJ\alpha_1^2-i\Omega. \label{ceom}
\end{eqnarray} 
Expressing the complex amplitude as $\alpha_j=R_j\exp{i\phi_j}$ the classical equation of motion (Eq.(\ref{ceom})) can be separated into following amplitude and phase equations:
\begin{eqnarray}
\dot{R_1}&=&\left(\frac{\gamma_1}{2}-\gamma_2R_1^2\right)R_1+2JR_1R_2\sin{\Phi}, \nonumber\\
\dot{R_2}&=&\left(\frac{\gamma_1}{2}-\gamma_2R_2^2\right)R_2-JR_1^2\sin{\Phi}-\Omega\sin{\phi_2}, \nonumber\\
\dot{\phi_1}&=&-\Delta_1-2JR_2\cos{\Phi},\nonumber\\
\dot{\phi_2}&=&-\Delta_2-J\frac{R_1^2}{R_2}\cos{\Phi}-\frac{\Omega}{R_2}\cos{\phi_2},\label{apeom}
\end{eqnarray}
where $\Phi=\phi_2-2\phi_1$ is the phase difference.In the absence of the external drive term ($\Omega=0$) we previously investigated the dynamics of the system in Eq. (\ref{apeom}) in Ref~\cite{nis}. We showed that the symmetry of the quadratic coupling permits only $1:2$ phase-locking and give rise to multistablity. As a consequence of this, the stable stationary states of the system evolves along the same periodic orbit with same amplitude but opposite direction of rotation. For  $\Omega\neq0$, with frequency detunings $\Delta_1=0$ and $\Delta_2=0$ the Eq.(\ref{apeom}) admits stationary states 
\begin{equation}
\cos\Phi=0, ~~~ \cos\phi_2=0.
\end{equation}
This conditions yields four possible fixed point corresponding to the combination 
\begin{equation}
\sin\Phi=\pm1 ~~~ \sin\phi_2=\pm1.
\end{equation}
 The linear stability analysis reveal that only steady-state solution that satisfies, $\Phi=\pi/2$ and $\phi_2=3\pi/2$ are stable. Owing to the $\pi$-periodicity introduced by the quadratic coupling the first oscillator possesses two symmetry related phase locked states located at $\phi_1=\pi/2$ and $3\pi/2$. Consequently, the first oscillator exhibits a two-fold phase degeneracy corresponding to $2:1$ phase-locking at frequency $\omega_1=\omega_d$, whereas the driven oscillator possess $1:1$ phase-locked state with the external drive at second harmonic frequency $\omega_2=2\omega_d$. The corresponding relative phase $\Phi$ is locked at $\pi/2$ at frequency $\omega_2=2\omega_1$. The classical phase-locking states are in excellent agreement with the quantum phase distributions as presented in Sec. \ref{quant}, where phase distribution of first oscillator $P(\phi_1)$ exhibits two maximas at $\pi/2$ and $3\pi/2$, phase distribution of second oscillator $P(\phi_2)$ exhibits single maxima at $\phi_2=3\pi/2$ and the relative phase distribution possesses a single maxima at $\Phi=\pi/2$. Therefore, the classical equations capture the essential asymmetric phase-locking geometry observed in the quantum regime. 

\end{document}